\documentclass[graybox]{svmult}

\usepackage{type1cm}        
\usepackage{makeidx}         
\usepackage{graphicx}        
\usepackage{multicol}        
\usepackage[bottom]{footmisc}

\usepackage{newtxtext}       %
\usepackage[varvw]{newtxmath}       

\makeindex             

\begin{document}

\title*{High-Contrast Imaging of Forming Protoplanets: VLTs, JWST, and the Promise of ELT}
\author{Gabriele Cugno \orcidID{0000-0001-7255-3251} and\\ Michael R. Meyer\orcidID{0000-0003-1227-3084}}
\institute{Gabriele Cugno \at Department of Astrophysics, University of Zurich, Winterthurerstrasse 190, 8057 Zurich, \email{gabriele.cugno@uzh.ch}
\and Michael Meyer \at Department for Astronomy, University of Michigan, 1085 S. University, 48109 Ann Arbor, Michigan, USA \email{mrmeyer@umich.edu}}
%
%
\maketitle


\abstract{Planet formation remains a fundamentally important yet poorly understood process. Protoplanetary disks, the birthplaces of planetary systems, exhibit a wide range of substructures that are increasingly interpreted as signatures of interactions with forming planets. However, the direct detection rate of protoplanets within these disks remains low, leaving critical gaps in our understanding of the physical mechanisms driving their formation and early evolution.
In this chapter, we review recent efforts by the high-contrast imaging community to directly observe forming protoplanets and their immediate environments. These observations aim to provide key constraints on thermal and accretion processes, planetary growth, and the formation of circumplanetary disks and satellite systems. We also propose a path forward for deriving observational estimates of the planet mass-to-radius ratio ($M_p/R_p$), a crucial parameter for distinguishing between competing formation models and understanding the thermal evolution of young planets. Finally, we highlight how upcoming instruments on the Extremely Large Telescope (ELT), with their unprecedented combination of high spatial and spectral resolution, will transform our ability to probe planet formation at the smallest and most critical scales.}



\section{Protoplanetary disks -- The birth place of planets}
\label{sec:disks_planets}

\begin{figure}
    \centering
    \includegraphics[width=\linewidth]{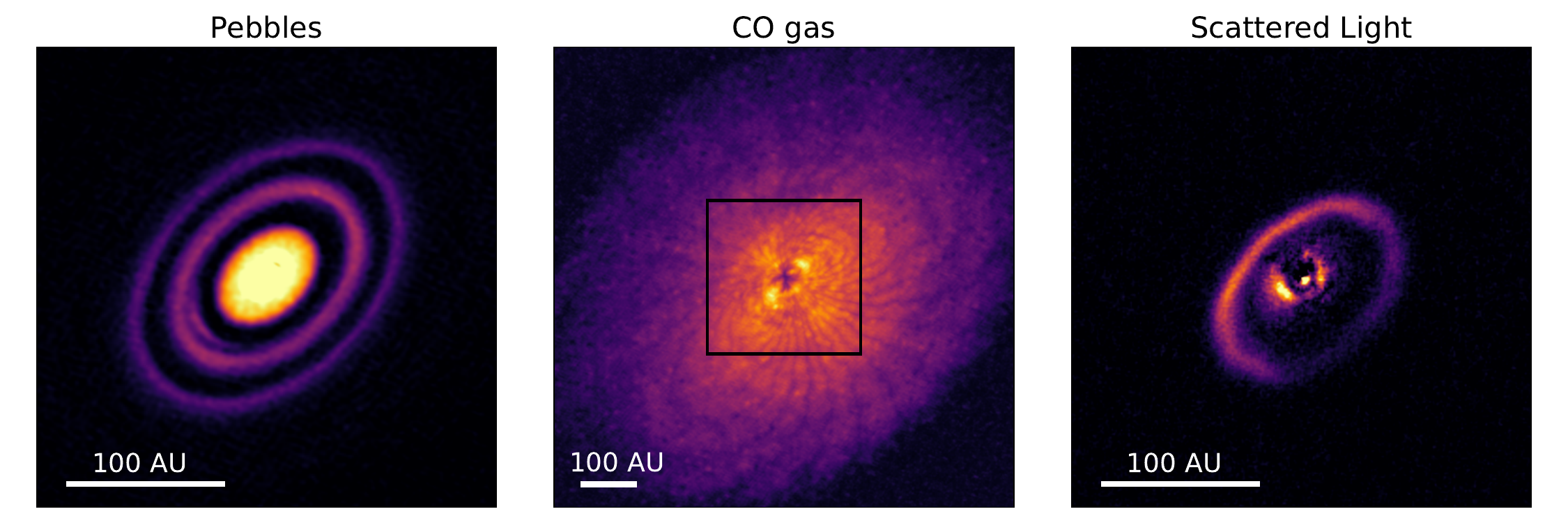}
    \caption{Disk surrounding the young star HD163296, as seen by ALMA in the continuum (left, see \cite{Andrews2018}) and CO gas (center, see \cite{Oberg2021_MAPS}), and by VLT/SPHERE in the $H$ band (right, see \cite{Ren2023}). Because the CO traces the entire extent of the disk, the central panel is shown on a different scale, while left and right panels report the same central disk region, which is highlighted in the central panel by a black square.}
    \label{fig:HD163}
\end{figure}

Over the past decade, significant progress has been made in the study of protoplanetary disks, driven by advances in observational techniques and instrumentation. Each observational method offers a unique window into different components of these complex environments. The Atacama Large Millimeter/submillimeter Array (ALMA) has revolutionized our understanding of disk structure by tracing millimeter-sized dust grains (e.g., \cite{Andrews2018}) and molecular gas distributions (e.g., \cite{Oberg2021_MAPS}) with unprecedented resolution. High-contrast imaging (HCI), primarily in the near-infrared (NIR), probes micron-sized dust at the surface layers (e.g., \cite{Benisty2023_pp7}), revealing scattered-light structures. The disk around HD163296 is shown in Fig.~\ref{fig:HD163} through these three tracers. Furthermore, mid-infrared (MIR) observations provide access to the chemistry and physical properties of the spatially unresolved inner disk (e.g., \cite{Henning2024}), where terrestrial planet formation is thought to occur. Together, these methods have greatly refined our understanding of disk structure and evolution.

\begin{figure}
    \centering
    \includegraphics[width=\linewidth]{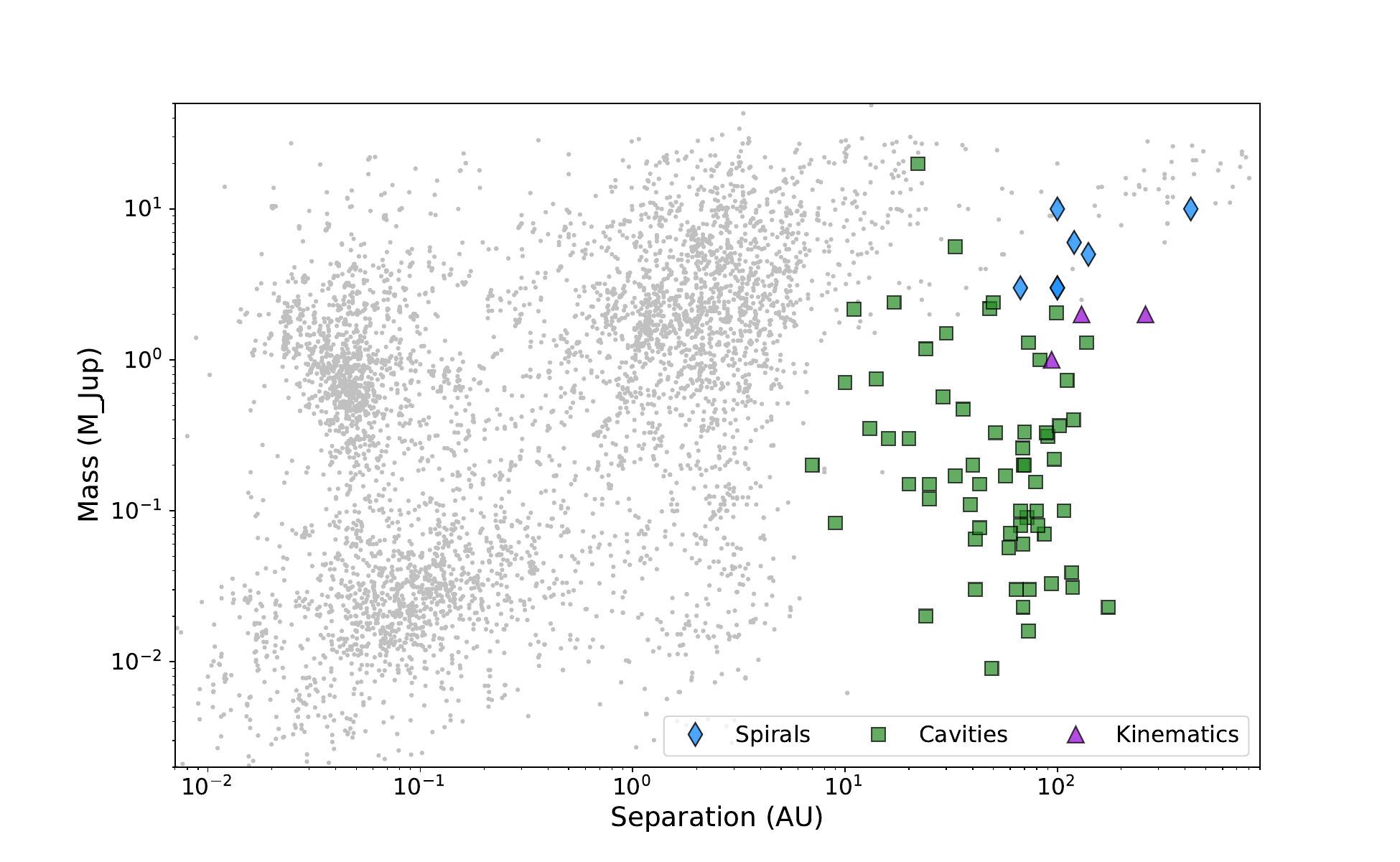}
    \caption{Mass versus semimajor axis of hypothesized planets assumed to reproduce the observed structures (gaps, spirals and kinematic perturbations), based on planet–disk interaction simulations. 
    The gray circles represent confirmed exoplanets as of April 2025 (https://exoplanetarchive.ipac.caltech.edu/). Protoplanet predictions are taken from \cite{Bae2023_pp7}. }
    \label{fig:pop}
\end{figure}

A key result emerging from these studies is that the disks surrounding young stars are highly structured, exhibiting gaps, rings, asymmetric spirals, and localized overdensities across all tracers, even at ages $<$ 1 Myr. These features often align with expectations from numerical simulations of planet-disk interactions, supporting the idea that embedded protoplanets shape their environments. Thus we are sensitive to a population of planets that can not currently be probed with other detection techniques (see Fig.~\ref{fig:pop}). However, alternative explanations exist, such as iceline-induced dust traps or magnetohydrodynamic effects, complicating the overall picture and interpretation (see \cite{Bae2023_pp7} for a comprehensive review).  The inferred properties of putative protoplanets are strongly dependent on modeling assumptions, and degeneracies in simulations introduce significant uncertainties.  This limits our ability to link specific structures to the presence of forming planets.

Beyond morphology, another promising approach to inferring the presence of forming protoplanets involves studying kinematic perturbations in the surrounding gas \cite{Pinte2023_pp7}. Massive protoplanets create localized deviations in the velocity field, often referred to as ``Doppler flips'' or ``kinks''. By analyzing these deviations, it is possible to infer the presence and mass of forming planets. This method provides a more direct tracer of gravitational interactions between protoplanets and their natal disks.

With mounting indirect evidence for an embedded population of protoplanets, the next natural step is to attempt to directly image these objects in the act of forming. Direct imaging has the potential to reveal not just the locations of protoplanets but also their physical and chemical properties, including temperature, luminosity (and therefore radius), atmospheric composition, as well as accretion activity. These observations would help bridging the gap between theoretical predictions and observations, and represent a fundamental breakthrough in linking the outcomes of planet formation with the initial conditions, a requirement for a predictive theory of planet formation.  

In this chapter we will first present the current state of observations of young protoplanets via high-contrast imaging, and describe the current efforts in characterizing the mature planet population on wide orbits. Then, we will present an approach that relies on as few as possible  underlying assumptions and would be key to empirically understand the initial conditions of gas giant planets and how they come into being. We conclude focusing on the major promises of the ELT, which is set to transform this field with its resolution, both angular and spectroscopic.

\section{What we know about protoplanets}
\label{subsec:direct_imaging_protoplanets}
Over the past two decades, tremendous effort has been devoted to the direct imaging of protoplanetary disks in the search for forming planets using a variety of instruments and observational techniques. Despite these efforts, the number of confirmed detections remains remarkably low. To date, the most compelling and robust case is the PDS70 system \cite[see Fig.~\ref{fig:PDS70}]{Keppler2018, Muller2018, Haffert2019}, which hosts two confirmed protoplanets (PDS70~b and c) and one candidate \cite[PDS70 d]{Christiaens2024, Hammond2025} embedded within a gapped circumstellar disk. In addition, a handful of intriguing candidates have emerged, such as AB~Aur~b and HD169142~b \cite{Currie2022, Hammond2023}, but these remain under active investigation and debate.

The low number of detections reflects not only the intrinsic difficulty of these observations, often contaminated by disk signals and complicated by post-processing artifacts, but also the complex and varied nature of the signals associated with forming planets. Protoplanets interact with their environment in multiple ways, each offering a potential observational tracer. For example, thermal radiation from the compact object, particularly in the near-infrared, can reveal the planet itself if it is hot and unobscured. Furthermore, accretion shocks onto the planet or its surrounding circumplanetary disk can produce emission lines that provides clues to the growth process. Finally, emission from circumplanetary material, either from warm dust in the mid-infrared or cold dust in the (sub-)millimeter, can also provide indirect but powerful evidence for angular momentum evolution of planet-forming material and constraints on satellite formation. In the next subsections, we will examine in more detail these three direct tracers of planet formation. For each, we will explore what observational signatures we aim to detect, what physical properties of the forming planets can be inferred from the data, and how ongoing observational programs can enhance our ability to uncover and characterize the planet formation process.

\subsection{Thermal emission from protoplanets}
\label{subsec:atmospheric_emission}

A key aspect to correctly interpret the evolution of mature systems is the initial conditions of protoplanets at the end of the formation process.
In particular, the amount of entropy retained by the planet is considered to be a crucial parameter to determine how planetary systems evolve over time (e.g., \cite{Spiegel2012, Mordasini2017, Marley2007}). If a lot of entropy is released during the gas accretion, then these planets will start colder and undergo a less significant contraction over the first 100 Myr of their lives. If accretion does not release a lot of energy, this is then retained by the system, and planets will start much hotter. Over the first phases of their lives, they will then cool down at a much higher rate, with cooling curves converging at about 300 Myr (e.g., \cite{Spiegel2012}). To distinguish between these scenarios, it is important to detect and study protoplanets during their formation phase. Indeed, while the bulk luminosity of different combinations of planet mass and specific entropy can be the same, the temperature (and thus radii) can distinguish between the two. Direct imaging of protoplanets is uniquely suited to provide these constraints. Furthermore, for these systems protoplanet masses could be obtained comparing planet-disk interaction numerical simulations with observations of disks. 

Moving to slightly older ages, the observations of young ($\lesssim30$~Myr) systems so far suggest that the known gas giants are not following a cold-start track (e.g., \cite{Balmer2025}). However, it is not yet clear if this is partially an observational bias --as we tend to detect warm- to hot planets since they are brighter-- or if we are really drawing meaningful conclusions on planet formation and evolution. The final release of Gaia data should help answer this question, as the astrometric mission does not rely on the planet brightness to infer the presence of massive companions to stars.

\begin{figure}
    \centering
    \includegraphics[width=0.5\linewidth]{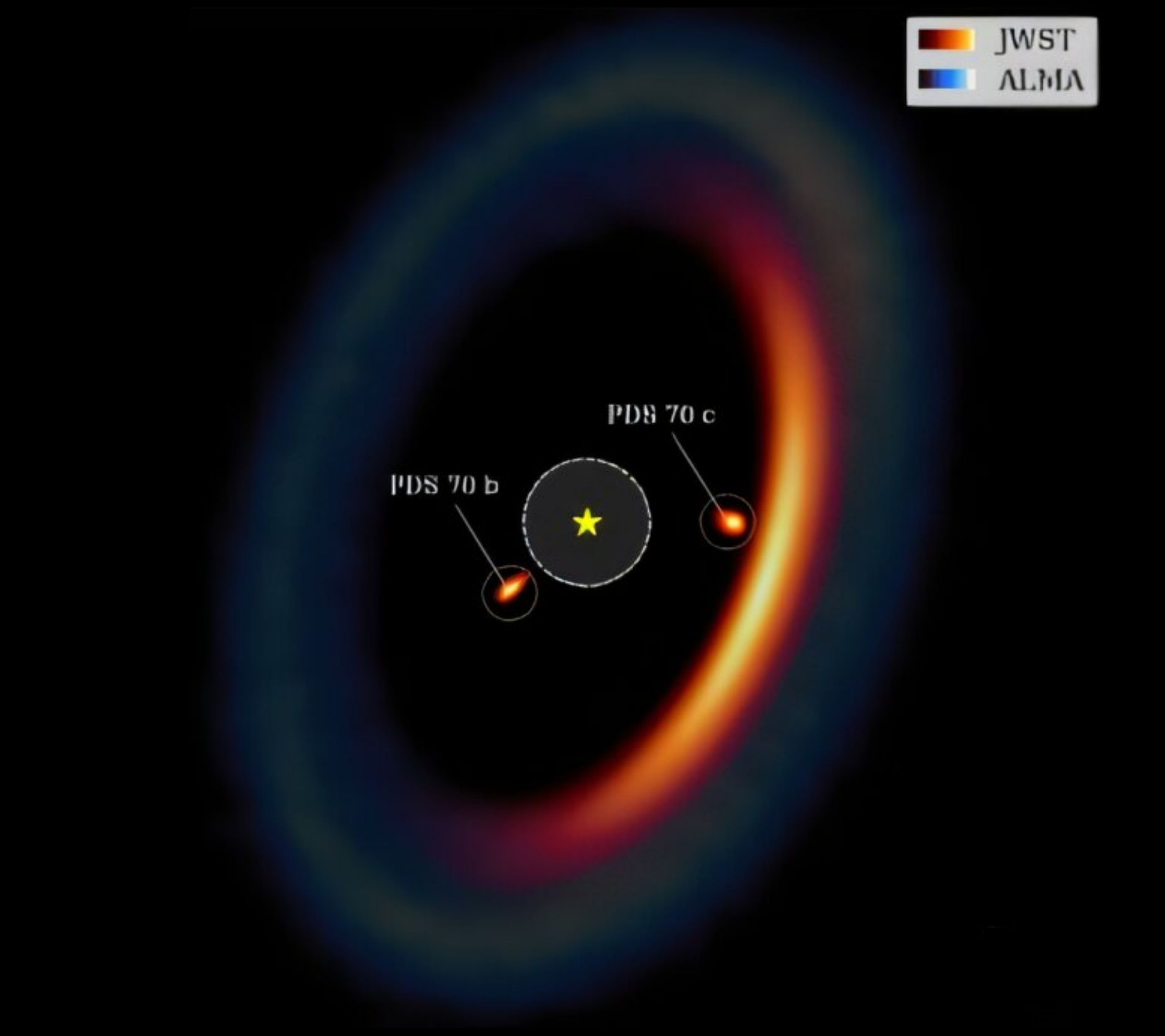}
    \caption{Composite image of the PDS70 system. The blue-scaled image reports the pebble ring as published in \cite{Benisty2021}, while the orange-colored image reports the reconstructed image obtained with JWST/NIRISS from \cite{Blakely2025}. The disk observed in the NIR is visible only on the inner rim of the near} side of the ring, where the stellar light scatters on the rim. Interior of the cavity, the two confirmed accreting protoplanets PDS70 b and c are detected at high significance. The central star and an inner disk detected at radio wavelengths have been masked.
    \label{fig:PDS70}
\end{figure}

Multiple HCI surveys focused on young stars with protoplanetary disks were conducted in the NIR with instruments like VLT/SPHERE, VLT/NaCo and Keck/NIRC2 \cite{Asensio-Torres2021, Cugno2023_ISPY, Wallack2024}. These resulted in a limited number of detections, like the PDS70 system (see Fig.~\ref{fig:PDS70}), with sensitivities to planet masses $\gtrsim$ few $M_J$ assuming hot-start models. These observations seem to suggest an inconsistency between the predictions from numerical simulations that reproduce disk observations and the actual planets sculpting protoplanetary disks. 
One solution to this apparent inconsistency is extinction from circumplanetary and circumstellar material, which could have significant impact on detectability \cite{Cugno2023_ISPY, Sanchis2020}. Furthermore, scattering from circumstellar disk material is brighter at NIR wavelengths, making the deetection of protoplanet signals much more difficult. These scenarios are supported by some recent tentative detections of protoplanet candidates, showcasing strongly extincted signals co-located with scattered light from the disk and/or planet environment \cite{Currie2022, Wagner2023, Zhou2023, Hammond2023}. 
To overcome the limitations associated with extinction and scattered light, in recent years there has been a push to search for protoplanets at longer wavelengths, where a reddened object would emit most of its emission. Longer wavelength observations may also benefit from the flux contribution of circumplanetary emission (see Section~\ref{sec:circumplanetary_disks}), making the detection of protoplanets easier. However, ground-based observations suffer from limitations related to the strong background emission coming from our atmosphere, the telescope and the instruments themselves. 

\begin{figure}
    \centering
    \includegraphics[width=\linewidth]{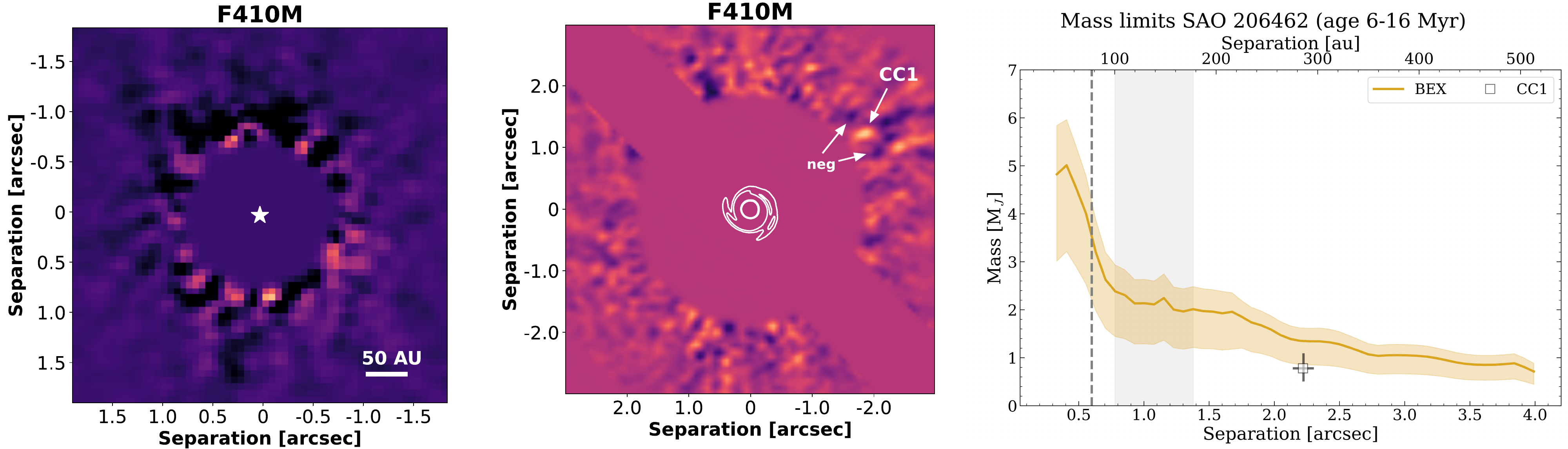}
    \caption{JWST/NIRCam results on the SAO206462 system presented in \cite{Cugno2024_NIRCam}. The left panel shows the residuals in the F410M filter in the area of the disk where an orbiting gas giant was predicted. The central panel shows a faint, yet robust, detection of a companion candidate (CC1) at large separation, best visible when a large mask is applied. The central panel also shows the disk scattered light signal from \cite{Ren2023}. The right panel reports the mass detection limit as a function of separation obtained using the warm-start models from \cite{Linder2019}. Planets above the limits could be excluded by the NIRcam data. These observations are in conflict with predictions from numerical simulations, which suggested a planet with $M\gtrsim5~M_J$ orbiting between $0.8-1.4$ arcsec (gray area). CC1 is also shown in the image. }
    \label{fig:SAO}
\end{figure}

The advent of JWST enabled deep observations at long wavelengths due to low background, overcoming the limitations of ground-based observations albeit at larger angular separation. Several programs in the first cycles, both part of Guaranteed Time Observations (GTO) or General Observers (GO) programs, targeted young disks showcasing significant structures with the specific intent of detecting protoplanets. Some of these observations have already been published, mostly at $\lambda\sim4~\mu$m, without providing robust detections of point sources in disks and actually often providing limits significantly deeper that the current planet predictions. 
An emblematic example is SAO206462, a disk showcasing two prominent spirals (e.g., \cite{Stolker2016}) that are thought to be launched by protoplanets according to multiple numerical studies. The disk is not massive enough to undergo disk instability (see Disk Instability chapter from R: Helled for a more complete discussion of this process). Overall, results agree that the protoplanet responsible for launching the spirals should be $>5~M_J$ at $100-120$~AU \cite{FungDong2015, Bae2016, DongFung2017}. \cite{Cugno2024_NIRCam} used deep JWST/NIRCam observations to search for the protoplanet. The data revealed no signal consistent with the predictions (Fig.~\ref{fig:SAO}, left panel), and mass detection limits reach a sensitivity of $\sim2~M_J$ (Fig.~\ref{fig:SAO}, right panel) derived from the BEX isochrones \cite{Linder2019}. A companion candidate has been identified at much larger separations ($\sim300$~AU, see central panel of Fig.~\ref{fig:SAO}). Despite being consistent with the initial observations of spiral motion \cite{Xie2021}, more recent data spanning a longer baseline indicate that the spirals move together and the perturber responsible for launching them should be located around $66\pm3$~au \cite{Xie2024}. Similar non-detections were obtained by \cite{Wagner2024}, with insufficient sensitivity to confirm or reject the MWC758\,c candidate, and \cite{Mullin2024}, who tried to search for protoplanets in the iconic HL Tau system.
These results point toward three possible scenarios: (i) either the planets form very cold ($\lesssim600$~K), (ii) they are highly extincted, and/or (iii) the predictions are not correct. Indeed, a very cold start would explain many non-detections of predicted planets embedded in disks in the thermal continuum. In the second scenario, the planet is deeply embedded in the circumplanetary disk/envelope, and the high extinction is hindering its detection at NIR wavelengths. \cite{Cugno2024_NIRCam} estimated that a visual `ISM-like' extinction of $A_V^\mathrm{ISM}\approx 20$~ mag would be necessary to make sure that a $5~M_J$ perturber following the BEX models would not appear in the NIRCam data of SAO206462.

The PDS70 system (Fig.~\ref{fig:PDS70}) remains the most compelling example of directly imaged, actively forming planets carving out a deep gap within a protoplanetary disk{\footnote{We note that after the writing of this chapter, a new protoplanetary disk with at least one forming embedded protoplanet (WISPIT~2b) has been found \cite{vanCapelleveen2025, Close2025_wispit, Facchini2026}.}}. Both PDS70\,b and c have been detected in the near-infrared, and significant efforts have been made to characterize their thermal emission. Spectroscopic observations with GRAVITY on the VLTI have provided tentative evidence of molecular features in the atmosphere of PDS70\,b \cite{Wang2021}, hinting at the presence of water and possibly other species, although the signal remains at the edge of detectability. Earlier attempts at atmospheric characterization using medium-resolution spectroscopy with SINFONI ($R\sim3500$) did not reveal any clear features \cite{Cugno2021}, highlighting the difficulty of such observations in the presence of surrounding disk material. More recently, higher resolution ($R\sim30'000$) KPIC data has yielded another tentative detection of molecular absorption, further motivating continued efforts to probe the planet’s atmosphere \cite{Hsu2024}. In parallel, GRAVITY’s long-term astrometric monitoring has begun to constrain the dynamical masses of the two planets \cite{Wang2021, Trevascus2025}. While the uncertainties are still substantial, the current estimates, under the assumption that the observed emission originates from the planetary photosphere, are inconsistent with cold-start evolutionary models. This suggests that PDS70~b and c likely formed with higher initial entropies, placing them closer to hot- or warm-start formation scenarios.

An important upcoming development for thermal imaging of protoplanets is SAXO+, the planned upgrade to the SPHERE extreme adaptive optics system at the VLT \cite{Vidal2022, Goulas2024}. SAXO+ will enhance the current SAXO (SPHERE AO for eXoplanet Observation) by improving wavefront sensing sensitivity and increasing correction speed, which will significantly boost achievable contrast at small angular separations. These upgrades are expected to reduce residual speckle noise and improve stability, allowing the inspection of the innermost region of protoplanetary disks, and in particular their central cavities.

\subsection{Accretion}
\label{subsec:accretion}

\begin{figure}
    \centering
    \includegraphics[width=0.6\linewidth]{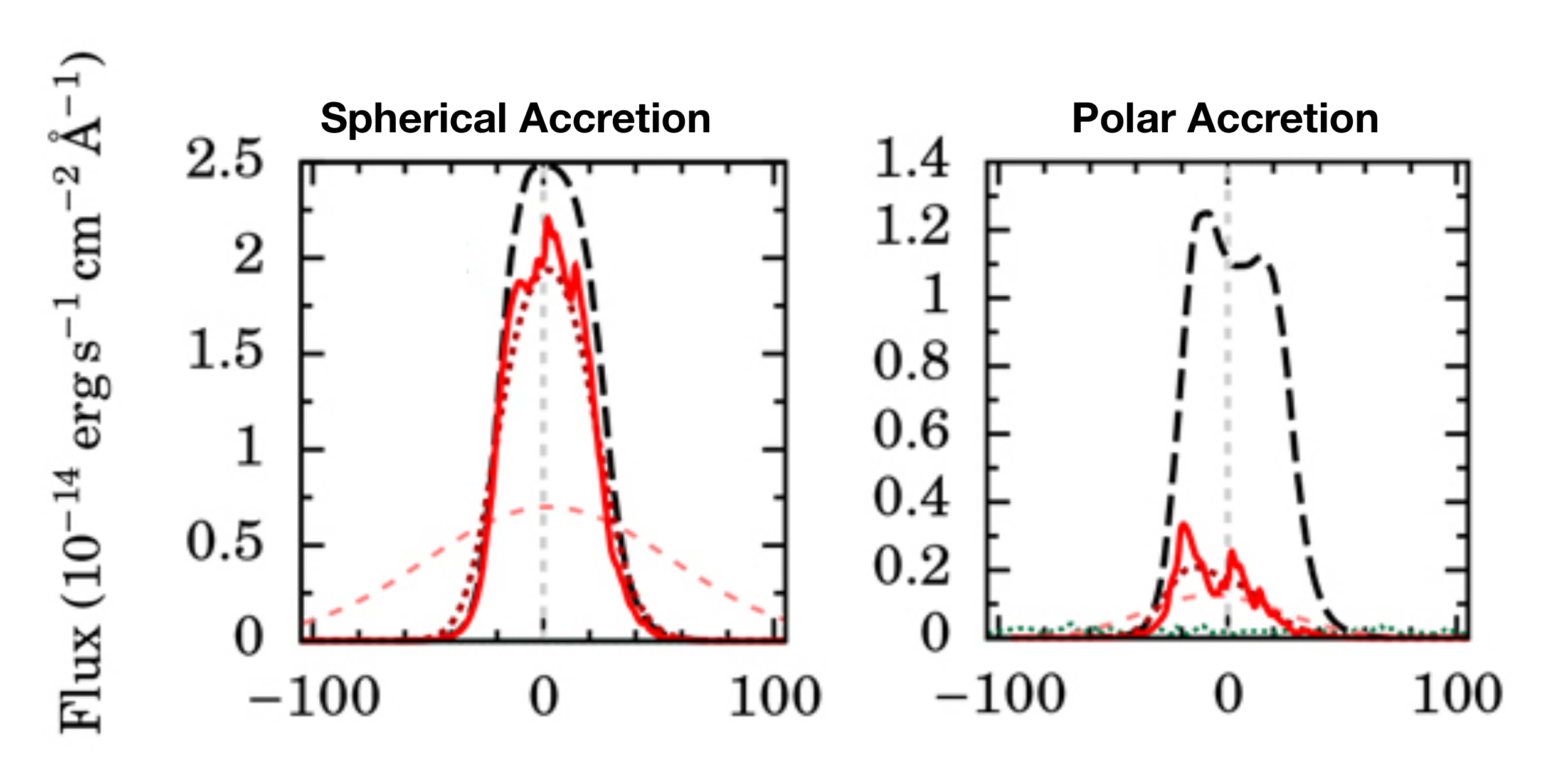}
    \caption{Expected H$\alpha$ line emission profiles (red) obtained from \cite{Marleau2023} for two spherical accretion geometries, spherical and polar accretion, after accounting for the extinction from the accretion flow. The dashed black line shows the profile before extinction is accounted for. The profiles are for $M_p = 2~M_J$ and $\dot{M} = 3\times10^{-5}~M_J$/yr. Obtaining high-resolution line profiles will enable the study of the accretion flow geometry. }
    \label{fig:Halpha}
\end{figure}

Studying the accretion of protoplanets through emission lines allows to understand how, when and at what rate the surrounding material is accreted with significant implications for the planet evolution and its final atmospheric properties. In particular, the line fluxes and their ratios can be used to derive the planet mass accretion rate. In addition, high resolution measurements of the line profile will strongly constrain the accretion geometry \cite[see Fig.~\ref{fig:Halpha}]{Marleau2023}, thus determining the accretion mechanism (magnetospheric accretion, boundary layer accretion, \cite{Thanathibodee2019, Owen2016}). 

H$\alpha$ ($\lambda\sim0.656~\mu$m) is expected to be the brightest line associated with protoplanets accretion processes. Hence, most initial efforts to search for accreting protoplanets focused on searches for localized H$\alpha$ emission using a technique called Spectral Differential Imaging (SDI; \cite{Close2014}). This approach relies on the instrument being equipped with two cameras observing simultaneously in two adjacent filters: one centered on the H$\alpha$ line, while the other in the nearby continuum. Given that protoplanets are relatively cold and have negligible thermal emission at optical wavelengths, the continuum image does not include any planet contribution, and can be used to model and subtract the stellar PSF from the H$\alpha$ image. The setup to employ SDI in H$\alpha$ is now available on multiple instruments (VLT/SPHERE, LCO/MagAO-X, Subaru/VAMPIRES) and has been used to target many circumstellar disks. Indeed, multiple surveys have been run with this technique, the first one being \cite{Cugno2019}, in which six stars hosting disks showcasing strong signposts of planet-disk interactions were inspected. Despite the known M-star companion to HD142527 \cite{Close2014}, no other companion was detected. Subsequently, additional efforts conducted both with SPHERE and MagAO were presented in \cite{Zurlo2020, Huelamo2022, Follette2023}, for a total of 23 stars surveyed. Despite the efforts, the detection yield remains low, with convincing detections in the planetary-mass regime only for the PDS70 planets \cite{Wagner2018, Close2025}. Additionally, more than one candidate was claimed in the literature, but confirmation is still pending (LkCa15~b, AB~Aur~b, \cite{Sallum2015, Currie2022}).

Motivated by the robust detections of PDS70\,b and c \cite{Haffert2019}, the community is also using Integral Field Units (IFUs) like VLT/MUSE to detect accreting protoplanets at H$\alpha$. These observations provide the significant advantage of higher spectral resolution, which helps in (i) increasing the SNR of the planet at high contrast, minimizing the continuum contribution from the star compared to a broader filter and (ii) spectrally distinguishing the planet emission thanks to slight offsets in radial velocity. Despite these advantages, searches for protoplanets conducted with VLT/MUSE have not increased the number of detections \cite{Xie2020}. 

Finally, HST can also provide high-contrast observations at H$\alpha$, even though these are limited in spatial resolution given the relatively small mirror of the space observatory compared to ground-based telescopes. The main advantage of HST observations is its capability to observe in the ultraviolet and detect the UV continuum \cite{Zhou2021}, hence directly probing $L_\mathrm{acc}$, which is usually derived from the line luminosity using a $L_\mathrm{line}-L_\mathrm{acc}$ relationship (untested in the planetary mass regime). Hence, measuring $L_\mathrm{acc}$ directly from UV observations helps in removing a rather uncertain step from the $\dot{M}_\mathrm{acc}$ derivation.  

Recently, variability in the H$\alpha$ emission of the PDS70 planets has been measured, with planet c appearing fainter in 2024 than in previous years \cite{Close2025, Zhou2025}. The variability has been measured with multiple instruments, confirming it is real. Additionally, \cite{Demars2023} recently inspected variability of the Pa$\beta$ line in the two accreting planetary mass companions GQ Lup B and GSC 06214-00210 b, finding variability on multiple timescales. Continued monitoring of these systems and of any other detected protoplanet in the coming years will be key to understand on what timescales does this variability occur, what is causing it and what are the implications for planet formation. One certain implication seems to be that it might be necessary to revisit targets multiple times when searching for accretion signatures, as it is possible that a non-detection is due to a period of quiescence in the protoplanet accretion processes. 

Until recently, extinction remained one of the main unknown effect potentially hindering the detection of protoplanets, as it is poorly characterized in protoplanetary disks and is usually expected to have a more significant impact at shorter wavelengths \cite{Cugno2019}. For example, \cite{Hashimoto2020} used the non-detection of H$\beta$ emission from PDS70~b and c to constrain the post-shock extinction to be $>2.0$~mag, assuming an `ISM-like' extinction law. For this reasons, it is expected that moving to redder emission lines at longer wavelengths would help reducing the impact of extinction from circumstellar and circumplanetary material, despite these lines being intrinsically fainter (hydrogen recombination line ratios can be found in \cite{Aoyama2021} under the assumption of no extinction). Whether or not this is going to be the case will strongly depend on the dust properties in the different protoplanetary disks and circumplanetary environments. Among the accretion lines in the NIR, one could look for, there are Pa$\beta$ at $1.28~\mu$m (e.g., \cite{Uyama2021}), Pa$\alpha$ at $1.87~\mu$m, Br$\gamma$ at $2.16\mu$m (e.g., \cite{Christiaens2019}) and Br$\alpha$ at $4.05~\mu$m. Note that Pa$\alpha$ and Br$\alpha$ can be observed with specific filters of the NIRCam instrument onboard JWST: \cite{Cugno2024_NIRCam} explored highly extincted scenarios for SAO206462, finding that a moderatly accreting protoplanet with $\dot{M} = 10^{-6}~M_J$~yr$^{-1}$ would have been detected through a medium with $A_V^\mathrm{ISM}=100$~mag. Overall, the search for emission lines at longer wavelengths did not provide detections, neither in the PDS70 system \cite{Christiaens2019, Uyama2021}, nor in other systems (e.,g. \cite{Cugno2024_NIRCam, Wagner2024}).

These studies assumed that extinction from circumstellar material behaves in the same way as in the ISM. However, \cite{Cugno2025} demonstrated for the first time that this is likely not the case. By performing transmission spectrophotometry on a background star shining through the disk gap in the AS 209 system, they empirically estimated the extinction affecting forming protoplanets, finding that it can reach several magnitudes even at $\sim4~\mu$m. Indeed, their analysis revealed evidence for a grey component of the extinction, which would lead to significantly stronger attenuation at longer wavelengths compared to ISM. This grey component is likely associated with dust grains larger than those in the ISM, produced through grain growth within the protoplanetary disk. 


To detected fainter signals at H$\alpha$, there is currently a push for developing new technology to improve AO correction at shorter wavelengths. This effort is mostly driven by the desire to use ground-based telescopes to detect planets in reflected light, but it can be critical to study planet formation as well. Tackling this problem  is the goal of the Extreme AO system MagAO-X \cite{Males2022}. Over the last three years, multiple upgrades were able to provide an improvement in the AO correction, significantly increasing the Strehl ratio of the observations \cite{Close2025}. Two convincing examples on this regard are \cite{Cugno2023_AS209}, where for the first time ground-based high-contrast observations in the optical were more sensitive than space-based HST observations at every separation, and \cite{Close2025}, where the improvement between the different epochs (and the MagAO data on the same target reported in \cite{Wagner2018}) is evident. These efforts in developing the technology to provide high-strehl diffraction-limited images at optical wavelengths will be instrumental to detect a larger sample of protoplanets, both with the currently available instrument and in the future with the ELTs. 

Another major technological advancement is the RISTRETTO instrument, led by University of Geneva \cite{Lovis2022}. RISTRETTO is a 7 spaxel high resolution integral field unit equipped with a coronagraph and fed by an extreme-AO system. While its main science goal is to detect reflected light planets, its unique capabilities will also be used to spectrally resolve and measure line profiles of accreting protoplanets. Encoded in these profiles there is a large amount of information on which are the dominating processes during accretion onto protoplanets and what is the current mass accretion rate. For example, \cite{Ringqvist2023} observed the spectrum of the well separated planetary mass companion Delorme~1~AB~b ($M_p = 12-14~M_J$) at high resolution without support from an AO system and found that the accretion line profiles are explained by a combination of narrow and broad components with different velocity shifts, and their results hint towards magnetospheric accretion being the dominant accretion mechanism. With instruments like RISTRETTO, this type of analysis will be possible for embedded protoplanets closer to their stars, like PDS70b and c. Furthermore, it is thought that the line width is a better avenue to infer mass accretion rates for compact objects, rather than line intensities. RISTRETTO will be instrumental in obtaining these types of measurements.


\subsection{Circumplanetary disks}
\label{sec:circumplanetary_disks}

During the formation of giant planets, circumplanetary disks (CPDs) of gas and dust emerge naturally due to the conservation of angular momentum from infalling material. These disks serve as a key interface between the forming planet and its surroundings, playing a fundamental role in several aspects of planetary evolution. CPDs regulate the process of atmospheric accretion, influencing the energy budget of the young planet and shaping its long-term thermal evolution. Additionally, they govern the properties of the gas that is incorporated into the planet’s atmosphere, affecting its final composition. Understanding CPDs is therefore essential not only for constraining the formation mechanisms of giant planets but also for deciphering the origins of their satellite systems, which coalesce from the material within these disks.

Theoretical models predict that CPDs are bright in the MIR, a consequence of the thermal emission from dust heated by the accretion shock, the planet atmosphere and the surrounding environment. Simulations suggest that these disks can reach temperatures of several hundred kelvin, producing significant excess emission in the $10-20~\mu$m range \cite{Szulagyi2019, Choksi2023, Sun2024, Taylor2025}. For instance, a CPD with a characteristic temperature of 500~K and an emitting area comparable to a few Jupiter radii would radiate strongly at MIR wavelengths, making it detectable with state-of-the-art instruments. These predictions set the stage for observational searches aimed at identifying and characterizing CPDs with high-contrast imaging means.

Empirical evidence supporting these predictions has already begun to emerge. The first clear indication of MIR excess associated with a protoplanetary companion came from \cite{Stolker2020}, who reported excess emission around PDS70\,b, consistent with the presence of a CPD. This finding was later reinforced by JWST observations, with both NIRCam \cite{Christiaens2024} and NIRISS \cite{Blakely2025} confirming excess emission around both PDS70\,b and c. While contamination from the circumstellar disk complicates photometric extraction in the NIRCam data, the NIRISS reduction allows for a more robust simultaneous fit of both the planet and the circumstellar disk signal, strengthening the evidence for CPD-associated emission. 

Radio observations from ALMA provide additional insights into CPD properties. ALMA continuum detections can help constrain the extent of circumplanetary material, assuming that the observed emission is optically thick \cite{Rab2019, BalleringEisner2019}. To date, only the CPD surrounding PDS70~c has been confirmed through ALMA continuum observations \cite{Benisty2021}, while other searches have so far yielded null results \cite{Andrews2021}, suggesting that CPDs may generally be compact and faint at millimeter wavelengths. ALMA has also been used to search for CPDs through molecular line emission, with \cite{Bae2022} reporting a candidate detection in $^{13}$CO around AS209~b. However, the existence of this candidate remains unconfirmed, as direct imaging follow-up studies have yet to confirm the presence of a planetary companion \cite{Cugno2023_AS209, Cugno2023_ISPY, Asensio-Torres2021}.

Sitting between the spectral ranges accessed by JWST’s near-infrared instruments and ALMA’s millimeter wavelengths, JWST’s mid-infrared capabilities offer a unique opportunity to explore an uncharted observational window. The $10-15~\mu$m range, accessible through the MIRI instrument, is particularly promising for detecting forming planets embedded in dense circumplanetary material radiating in the MIR range. Although no MIRI datasets targeting protoplanets have been published thus far, upcoming observations promise to push the detection limits to unprecedented sensitivities, providing a crucial missing piece in the broader picture of planet formation.

\begin{figure}
    \centering
    \includegraphics[width=\linewidth]{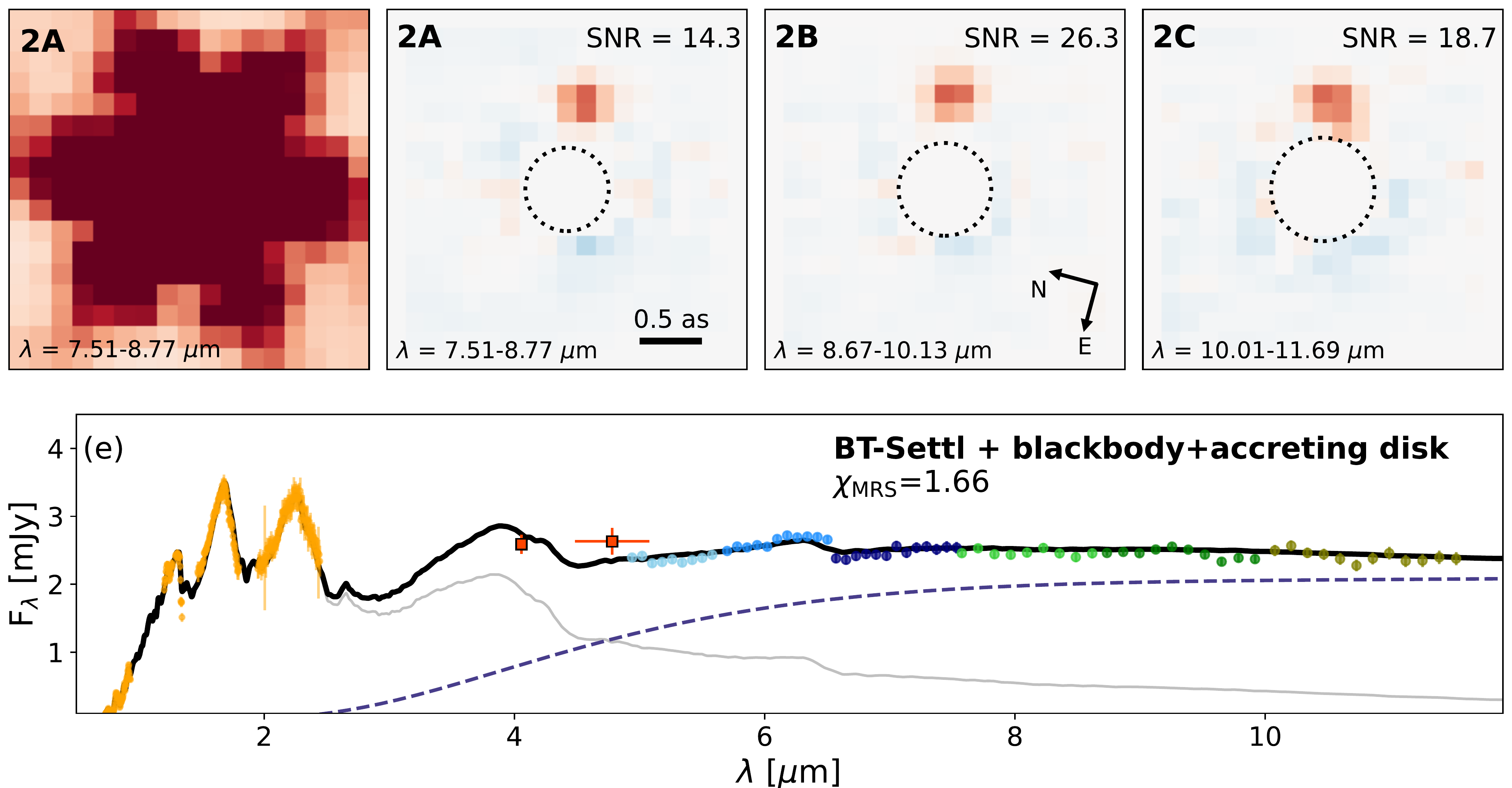}
    \caption{\textit{Top:} GQ Lup wavelength-combined images before (first column; only channel 2A shown) and after (second to fourth columns; channels between 2A and 2C) PSF subtraction, plotted with the same color scale. The corresponding wavelength range and S/N are shown in each panel. The central dashed circle represents the FWHM of the stellar PSF. With increasing wavelength, the size of the PSF increases and the companion separation falls to $\sim1.5~
    \lambda/D$. \textit{Bottom:} Best fit of the SED of GQ Lup B when using the BT-Settl model to describe the atmosphere and `blackbody+accreting disk' model to describe the disk emission \cite{Cugno2024_GQLup}. Orange data points represent MUSE and SINFONI data, red squares show VLT/NaCo photometries \cite{Stolker2021}, and circles beyond $5~\mu$m show the spectrum extracted from the MRS data. The gray line shows the atmospheric contribution from GQ Lup B, the dashed blue line shows the contribution from the disk, and the thick black line shows the overall SED model. Image adapted from \cite{Cugno2024_GQLup}.}
    \label{fig:CPD}
\end{figure}

Despite these promising approaches, angular resolution and contrast requirements will maintain the study of CPD proeprties (composition and spectral shape) very limited. Fortunately, we can access planetary mass companions at wider separations, which are easier to observe both in terms of resolution and contrast, and which are usually not embedded in the circumstellar disk. Some of these analogs show typical accretion signatures and MIR excesses, strongly indicating the presence of a CPD that can be studied in easier conditions. These systems provide a valuable comparison to CPDs embedded in protoplanetary disks, allowing to probe similar physical and chemical processes in more accessible settings. Recent work has already demonstrated the power of such analog studies. Notably, the first low-resolution spectra of a CPDs in the MIR, associated with GQ Lup B and Yses-1 b \cite{Cugno2024_GQLup, Hoch2025}, have opened new avenues for investigating protoplanetary environments at longer wavelengths (see Fig.~\ref{fig:CPD}). This discovery strengthens the motivation to search for CPDs around embedded forming planets around lower-mass companions using JWST, particularly with MIRI’s high-contrast imaging capabilities.

So far, CPD characterization has primarily relied on photometry and low-resolution spectroscopy, but JWST’s Medium Resolution Spectrograph (MRS) represents a potential game changer. With a spectral resolution of 1700–3500 in the MIR, the MRS enables detailed searches for molecular emission from circumplanetary environments. Detecting specific molecular features could provide crucial insights into how late-stage accretion processes shape the final atmospheric composition of giant planets, potentially revealing whether certain chemical signatures can be linked to the planet’s formation location within the disk \cite{Oberg2011}. {\cite{CugnoGrant2025} presented the first chemical characterization of the molecular gas coming from a CPD, by extracting the JWST spectrum from the planetary mass companion CT Cha b. They found that the CPD is very carbon-rich, in stark contrast to the chemistry found in the disk surrounding the host star, and implying that local processes likely drive rapid, divergent chemical evolution between the two disks of the same system.} Furthermore, the MRS covers multiple key dust features, making it possible to simultaneously constrain both the gas and dust properties of CPDs. Combining these data with ALMA’s constraints on millimeter dust emission \cite{Wu2022} and gas kinematics could ultimately provide a comprehensive picture of CPD structure, composition, and evolution, bringing us closer than ever to a complete understanding of how giant planets and their satellite systems take shape.

\subsection{Demographics of protoplanets and mature systems}

Tremendous progress has been made in the past decade in constraining the demographics of gas giant planets \cite{Gaudi2021}. We have significant constraints on the orbital distribution, mass function, and overall frequency of gas giant planets \cite{Vigan2021}. For example, the mean number of planets per star between $1-10~M_J$, integrated over all orbital separations is about 11.4\% \cite{Meyer2025}.  However, most of these planets are found between 1--10 AU \cite{Fulton2021} \cite{Meyer2025}, and if we restrict the frequency to those with orbital radii $>$ 30 AU, the frequency drops to 0.5\%.  Given the power--law nature of the mass function, if we were sensitive to planets down to Saturn's mass the frequency increases by a factor of $\times$ 2.  Caveats to these extrapolations are: a) we generally assume that the planet mass function does not depend on orbital separation (consistent with \cite{Fulton2021} down to 30 Earth masses) and \cite{Meyer2025}; and b) the fits to the gas giant population, which are robust, must break down somewhere below the mass of Saturn.  

While we have demographic data on gas giants over all separations, our knowledge of sub-Saturns is limited within a few AU, with the exception of planets down to 10 Earth masses from 1-10 AU from microlensing surveys, presumably around M dwarf hosts.  
Clearly ice giant and rocky planet formation is different than gas giant formation.  To date there are no constraints on the frequency of Neptune-sized planets beyond 10 AU, with attempted estimates inferred from disk structures observed around T Tauri stars \cite{Zhang2023}. Fortunately, large direct imaging surveys with JWST are underway which will address this question (Carter et al., in prep). \cite{Bogat2025} demonstrates that JWST can be sensitive down to 10 Earth masses for nearby young stars, meaning that if there are numerous ice giants beyond 10 AU around young stars in nearby moving groups, JWST will find them.  If JWST does not find these planets around stars whose disks are mostly dissipated, how can we explain the apparent discrepancy? One possibility is that $10-30$ Earth mass planets, expected to suffer rapid Type I migration, have relocated to the inner disk where they are detected by transit, radial velocity, and microlensing.  Alternatively, our evolutionary models could require adjustments, as they have not yet been validated with objects of known dynamical mass, age, and measured luminosities. Soon Gaia will be able to detect Neptune-mass planets around nearby stars, and such planets are in principle observable in reflected light with Roman and in thermal emission with METIS on the ELT.  With new observational constraints, better opacities for cold atmospheres needed for the lowest mass (and oldest) planets and additional model sophistication 
 hopefully we will be able to constrain the frequency of outer ice giants in a robust way.  

 This chapter is mostly devoted to high-contrast imaging observations of gas and ice giants in formation, in the presence of a gas--rich primordial disk whose properties are dictated as a natural consequence of the star formation process.  However, classical models of terrestrial planet formation inside the ice--line (about 3 AU) involve giant collisions of Mars-sized objects on self-stirred chaotic orbits after the dissipation of the primordial disk \cite{Wetherill1993}. As these planets grow, a 0.9 Earth mass protoplanet being hit by a 0.1 Earth mass protoplanet results in an Earth-sized body with a surface temperature of molten silicates. In fact such a scenario is thought to be the origin of the Earth--Moon system \cite{Benz1986}. \cite{Miller-Ricci2009} and \cite{Bonati2019} explored whether such a hot protoplanet collision afterglow could be observable \footnote{\cite{Mamajek2007} explored such an improbably scenario to explain the under luminosity of 2M12078 b}.  While such a hot protoplanet cools in about 10'000 years, a thick atmosphere could prolong the duration of a hot surface for millions of years. The catch is that the brightness temperature where the atmosphere becomes optically thick in such a scenario is only hundreds of degrees. Given the duration of the terrestrial planet forming epoch (about 100 Myr, \cite{Raymond2014}, but see also book chapters from S. Grimm and A. Kessler), the sensitivity of ELTs and future space-based infrared telescopes, and the number of nearby young stars in this age range, a search for hot protoplanet collision afterglows is feasible. Given the observed diversity in terrestrial planet bulk densities (see chapter by S. Grimm and A. Kessler), and the lack of compositional predictions from planet formation models (cf. \cite{Elser2012}), such observations would provide powerful constraints on planetary composition. 

\section{A path forward to unveil planet formation}

Building on the progress achieved with direct imaging of protoplanets and their environments, the field now faces the challenge of translating individual detections into robust physical constraints on planet formation. To do so, we must move beyond qualitative evidence and develop methods that can provide semi-empirical measurements of key parameters. A crucial quantity in understanding the physical processes that govern planet formation and the key parameters that dictate its outcome is the mass-to-radius ratio ($M_p/R_p$) of gas giant planets. 
It governs the contraction of protoplanets and provides an essential diagnostic for distinguishing between different evolutionary pathways. Specifically, hot-start formation models predict a relatively small $M_p/R_p$ ratio compared to cold-start models, making this parameter a powerful probe of planetary birth conditions. Finally, $M_p/R_p$ can be related to the temperature of the central object through structure equations. 

Several approaches can be employed in the coming years to determine $M_p/R_p$, each offering unique advantages and cross-validation opportunities. These methods are semi-empirical as they rely on as few assumptions as possible. Although not quantitative, they have revealed trends in large samples, and even common factors that can reveal separate classes of objects.\\



{\bf (i) Semi-empirical Measurements of Mass and Radius}

One strategy involves measuring the mass and radius independently and then deriving their ratio. The dynamical mass of a protoplanet can be determined through dynamical techniques, such as orbital fitting thanks to the exquisite astrometric precision provided by GRAVITY \cite{Trevascus2025}. In embedded environments, either hydrodynamical modeling or high-resolution observations of the broadening of the  CO line profile can be used. 
Indeed, the Doppler broadening of these lines is directly proportional to the square root of the planet’s mass, allowing for precise mass measurements (the measurement is degenerate with the disk inclination if the disk is not resolved). This technique is widely used to measure dynamical masses for stars, \cite{Simon2000}, and could be used for protoplanets too. 

For radius measurements, SED fitting in the NIR is the most promising option. By modeling the observed flux across multiple bands and providing estimates of effective temperature\footnote{If absorption lines are present in the planet's spectrum, the latter can be compared with the one of low-gravity objects of similar spectral types.} and luminosity, constraints on the planet’s radius can be derived. In order to undertake this approach, it will be essential to have a broad and well sampled $\lambda$ coverage to model the multiple components that are contributing to a protoplanet's SED (accretion shock continuum, central compact object, CPD). 
Additionally, extinction from surrounding material remains a significant challenge, as dust in the circumstellar disk can obscure or redden the planet’s intrinsic emission, biasing the radius measurement. Addressing this issue requires sophisticated modeling of dust properties, coupled with observational constraints potentially coming from background sources shining through the disk material \cite{Cugno2025}. \\

{\bf (ii) Using Accretion Diagnostics}

An alternative and complementary approach involves leveraging multiple accretion lines to infer both the extinction affecting the fluxes and the accretion luminosity. This method capitalizes on the fact that line fluxes at different wavelengths are affected differently by extinction, allowing for a self-consistent determination of the level of extinction if the wavelength-dependence is known \cite{Cugno2025}. Furthermore, line profiles, such as those provided by METIS observations of the Br-$\alpha$ line, offer a robust measurement of the mass accretion rate onto the planet.

The mass accretion rate ($\dot{M}_\mathrm{acc}$) is related to the accretion luminosity ($L_\mathrm{acc}$) through Eq.~\ref{eq:acc}:

\begin{equation}
M_\mathrm{acc} = \left(1 - \frac{R_p}{R_\mathrm{in}}\right)^{-1} \cdot \frac{L_\mathrm{acc} R_p}{G M_p}
\label{eq:acc}
\end{equation}

where $R_\mathrm{in}\gg R_p$ is the starting radius of the gas infall 
and usually has a weak dependence on the equation (in most cases so far $R_\mathrm{in} = 5R_p$ has been used). 
Hence, by independently deriving $M_\mathrm{acc}$ (from line profiles) and $L_\mathrm{acc}$ (from extinction-corrected emission line intensities), one can solve for $M_p/R_p$. Given accretion processes are known to be variable \cite{Zhou2025, Close2025}, it is essential that all these measurements occur as simultaneously as possible. If this is not the case, it is possible that values obtained at different epochs with planets accreting at different rates bias the $M_p/R_p$ measurement, leading to wrong interpretations and conclusions. This value can then be compared to the one obtained through method (i), validating the measurement. \\

To place planet formation in a broader context, these methodologies must be applied systematically across multiple forming protoplanets and young planetary systems. By expanding the sample size, it will be possible to investigate whether there exist distinct classes of forming planets with different evolutionary pathways. By combining dynamical mass measurements, extinction-corrected radius estimates, and accretion diagnostics, we can rigorously test formation models and refine our understanding of how gas giant planets grow and evolve. The next generation of (upgraded) observational facilities, JWST, ALMA, ELT/METIS, and GRAVITY+, will be instrumental in advancing this endeavor, bringing us closer than ever to a comprehensive picture of planet formation.

\section{The next eye in the sky: the Extremely Large Telescopes}

This multi-wavelength, multi-tracer approach is essential, as each observational window is sensitive to different stages and components of the planet formation environment. Yet, despite the diversity of tools available, the challenge remains formidable: many planets may be too deeply embedded, too faint, or too transient to be detected and studied easily. The next years of JWST observations will be crucial in order to {\it systematically} probe the most promising protoplanetary disks both in the NIR and MIR. This is an essential step to identify the best strategy for observing forming planets embedded within their natal disks before the ELT will become operational. With this information available, the enormous potential of the ELT instruments can be unleashed to study forming planets and their environments. 
As concluding remarks, we provide some of the most exciting science questions that will be addressed by the new instruments.

\begin{enumerate}
    \item Thanks to its improved angular resolution, HARMONI and METIS will enable the search for protoplanets at high contrast much closer to the water snowlines and in the cavities of transition disks, where the peak of the giant planets population is located ($3-10$~AU, \cite{Fernandes2019}). Thanks to the multiple filters available, they will study the colors of protoplanets, building their SEDs to constrain the luminosity, and hence temperature and size, of protoplanets and their CPDs.

    \item Using the high-resolution ($R\sim100'000)$ IFU mode of METIS, it will be possible to obtain the spectrum of the protoplanet and at the same time kinematically characterize the warm dust and gas directly sculpted by it, in particular the v(1-0)P08 CO line \cite[see Fig.~\ref{fig:METIS_CPD}]{Oberg2023}. This will provide strongly needed constraints on planet-disk interaction processes, and will enable the study of how the material flows from the circumstellar disk onto the protoplanet (through the circumplanetary disk). Furthermore, this approach will enable the measurement of protoplanet's dynamical masses through extensive numerical modeling.

\begin{figure}
    \centering
    \includegraphics[width=\linewidth]{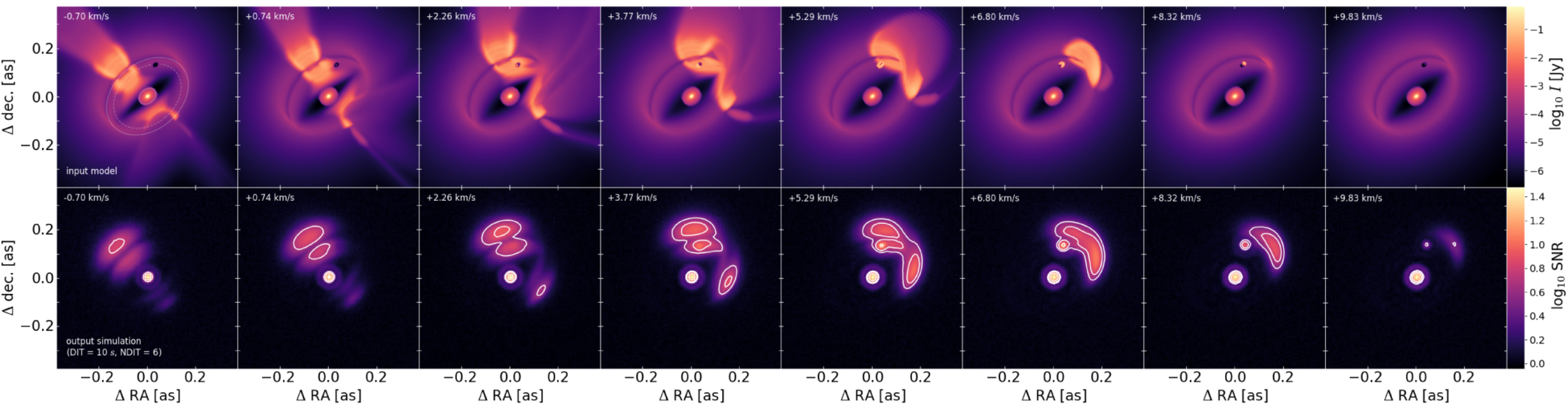}
    \caption{Synthetic channel maps of the v(1-0)P08 line (4.73587 µm) of the HD 100546 CSD+CPD model for the high external irradiation case. The CPD outer radius $R_\mathrm{out} = R_\mathrm{Hill}$ (1.62 au). The model used as input for the METIS simulation is in the top row. The corresponding simulated METIS observation panels represent six detector integrations of 10 s each in the bottom row of every subfigure. The white contour lines represent the signal-to-noise ratio values  of 3, 5, and 10. The velocity offset is relative to the stellar reference frame. Figure taken from \cite{Oberg2023}.}
    \label{fig:METIS_CPD}
\end{figure}

    \item ANDES and METIS will expand the discovery space in the field of accreting protoplanets. In particular, the high-resolution capabilities could be used to obtain the multiple line profiles for accreting protoplanets, extracting crucial information to constrain the thermal and dynamical structure of the accretion flow (see Fig.~\ref{fig:Halpha}). 
    \item A particularly exciting opportunity will come from the optical and near-infrared spectrograph HARMONI on the ELT, which will operate at wavelengths as short as $0.8~\mu$m. At this wavelength, CPDs could be detected in scattered light and even marginally resolved. Taking the PDS70c protoplanet as an example, we can estimate the spatial extent of its CPD by assuming it extends to one-third of the planet’s Hill radius. With a projected separation of 34~AU and an estimated mass of 2~$M_\mathrm{Jup}$, the resulting Hill radius is roughly 3.2~AU, yielding a CPD radius of about $\approx1$~AU. At the distance of PDS70 (113~pc), this translates to an angular diameter of $\sim9.3$~mas, well matched to the expected diffraction limit of HARMONI at $0.8~\mu$m ($\sim5.2$~mas). This implies that the CPD could be marginally resolved in reflected light, providing unprecedented spatial information on its geometry and substructure, and another direct constraint on the planet mass through $R_\mathrm{Hill}$. 
    \item METIS will obtain CO line profiles for protoplanets and their surrounding gas, measuring their masses dynamically. This step is fundamental to test and anchor evolutionary models at very young ages.
    \item 
    One key limitation of METIS in studying CPD emission will be its restricted wavelength coverage, imposed by the telluric absorption that renders large portions of the mid-infrared inaccessible from the ground. For spectroscopy studies at long wavelngths ($\lambda\gtrsim10~\mu$m), METIS will rely on a low-resolution ($R\sim400$) slit spectrograph in the $N$ band, making detailed molecular characterization of CPDs challenging at the very least. Nevertheless, the $N$-band spectrograph will cover the $10~\mu$m silicate feature, offering valuable insights into the dust grain population and mineralogy of CPDs.

\end{enumerate}
 
Swiss research groups have played a significant role in advancing the direct imaging of forming protoplanets, contributing to the development of major instruments (e.g., SPHERE including SAXO+, ERIS and RISTRETTO at the VLT, MIRI onboard JWST) and observing programs (e.g., SPHERE-SHINE, NaCo-ISPY). The significance of these efforts is reflected in the participation of Swiss institutions to the METIS and ANDES consortia, where unveiling planet formation is one of the  primary science goals. Future observing programs will provide the critical insights outlined above, bringing us closer to a comprehensive understanding of the processes that shape planetary systems. With the established expertise and strong international collaborations, Swiss researchers are well positioned to play a leading role in this next chapter of discovery.

\bibliographystyle{spmpsci}
\bibliography{main}

\end{document}